\documentclass[epj]{svjour}
\usepackage[english]{babel}
\usepackage{lmodern}
\usepackage[latin9]{inputenc}
\usepackage{endnotes}
\usepackage{amssymb,amsmath,amsfonts}
\usepackage{textcomp}
\usepackage{enumerate}

\usepackage{graphicx}% Include figure files
\usepackage{dcolumn}% Align table columns on decimal point
\usepackage{bm}% bold math
\usepackage{xcolor}

\begin{document}

\title{\boldmath Playing with Casimir in the vacuum sandbox}

\author{S. Kauffman\inst{1,2} \and S. Succi\inst{3,4,5} \and A. Tiribocchi\inst{3,4} \and P. G. Tello\inst{6}}

\institute{The Institute for System Biology, Seattle (WA), USA \and Professor Emeritus, Biochemistry and Biophysics, The University of Pennsylvania, (PA), USA \and Center for Life Nano Science@La Sapienza, Istituto Italiano di Tecnologia, 00161 Roma, Italy \and Istituto per le Applicazioni del Calcolo CNR, via dei Taurini 19, Rome, Italy \and Institute for Applied Computational Science, John A. Paulson School of Engineering and Applied Sciences, Harvard University, Cambridge, USA \and CERN, Geneva, Switzerland}

\date{\today}

\abstract{The Casimir effect continues to be a subject of discussion regarding its relationship, or the lack of it, with the vacuum energy of fluctuating quantum fields. In this note, we propose a {\it Gedankenexperiment} considering an imaginary process similar to a vacuum fluctuation in a typical static Casimir set up.  The thought experiment leads to intriguing conclusions regarding the minimum distance between the plates when approaching the Planck scale. More specifically, it is found that distance between the plates cannot reach a value below $(L/L_P)^{2/3}$ Planck lengths, being $L_P$ the Planck length and $L$ the typical lateral extension of the plates. Additional findings allow the conclusion that the approach between the two plates towards this minimum separation distance is asymptotic.}
\maketitle

\section{Introduction and aim}

The physical vacuum seems to be a ``busy place'' described as ``foamy'' by Wheeler \cite{Wheeler}, and one of the most intriguing phenomena linked to it is the Casimir effect \cite{Casimir}. In this note, we propose a {\it Gedankenexperiment} based on the Casimir effect by following the tradition of speculating for the playful purpose of thinking through its consequences. This {\it Gedankenexperiment} is suggested with the caveat that the relation of the Casimir effect with the vacuum energy of fluctuating quantum fields is still open to debate at the time of this writing \cite{Jaffe}.

\section{Gedankenexperiment setup}

The starting point is the typical configuration of the static Casimir effect: two uncharged, flat and perfectly conductive parallel plates in the vacuum, placed at a very small distance apart $d$ (typically in the order of microns or even a few nanometers). 

%\begin{figure}
%\includegraphics[width=1.0\linewidth]{fig1.pdf}
%\caption{{\it Gedankenexperiment} proposed in the text. In a typical Casimir setup, one of the virtual modes of the vacuum is ``removed'' with the consequent shrinkage of the initial distance, $d$, within the plates to $d'<d$.}
%\label{fig1}
%\end{figure}

Accordingly, the energy $E$ confined between the plates can be expressed as:
\begin{equation}\label{eq1}
E=-C\frac{\hbar cL^2}{d^3}.
\end{equation}

Here $C=\pi/24$, $c$ is the speed of light, $L$ is the lateral extension of the plates and the minus sign indicates a negative energy, corresponding to attraction between the plates. The {\it Gedankenexperiment} assumes the occurrence of a vacuum fluctuation between the plates (quantized by the value of $d$) subject to the energy-time relationship.  The Casimir plates, initially separated at a distance $d$, will be therefore momentarily brought together to a smaller distance $d'$, due to the fluctuation with energy $E_n=\hbar\omega_n$.  Accordingly, considering such fluctuation, the variation $\Delta E$ of energy between the plates will be given by the difference between the initial value and the final one, with the former is greater than the latter, since $C>0$ and $d'<d$. Thus, it is possible to write:
\begin{equation}\label{eq2}
\Delta E=\hbar\omega_n=C\frac{\hbar cL^2}{d'^3}-C\frac{\hbar cL^2}{d^3}.
\end{equation}

Let us consider now a standard interpretation of the energy-time indeterminacy relation which tells, in ordinary language, that if an amount of energy $\Delta E$ is ``borrowed'' from the vacuum, it must be ``returned'' quickly enough within a time interval $\Delta t$. In the context of this {\it Gendankenexperiment}, $\Delta t$ would correspond the fluctuation lifetime. Mathematically, it is expressed by the energy-time uncertainty relation $\Delta E\Delta t\ge \hbar/2$. Accordingly, $\Delta t$ is
\begin{equation}
  \Delta t \ge \frac{\hbar}{2\Delta E}.
 \end{equation}
Taking into consideration Eq.(\ref{eq2}), it is possible to write
\begin{equation}\label{eq4}
  \Delta t\ge\frac{\hbar}{2\Delta E}=\frac{1}{2\omega_n}.
\end{equation}
In this {\it Gedankenexperiment}, the boundary conditions assumed for the quantized mode between the plates allow $\omega_n$ to be expressed, for some integer number $n$, as:
\begin{equation}\label{eq5}
  \omega_n=\frac{2\pi nc}{d}.
 \end{equation}
Substituting Eq.(\ref{eq5}) into Eq.(\ref{eq4}) leads to the approximation
\begin{equation}\label{eq6}
\Delta t\sim \frac{d}{nc}.  
\end{equation}

\begin{figure}
\includegraphics[width=1.0\linewidth]{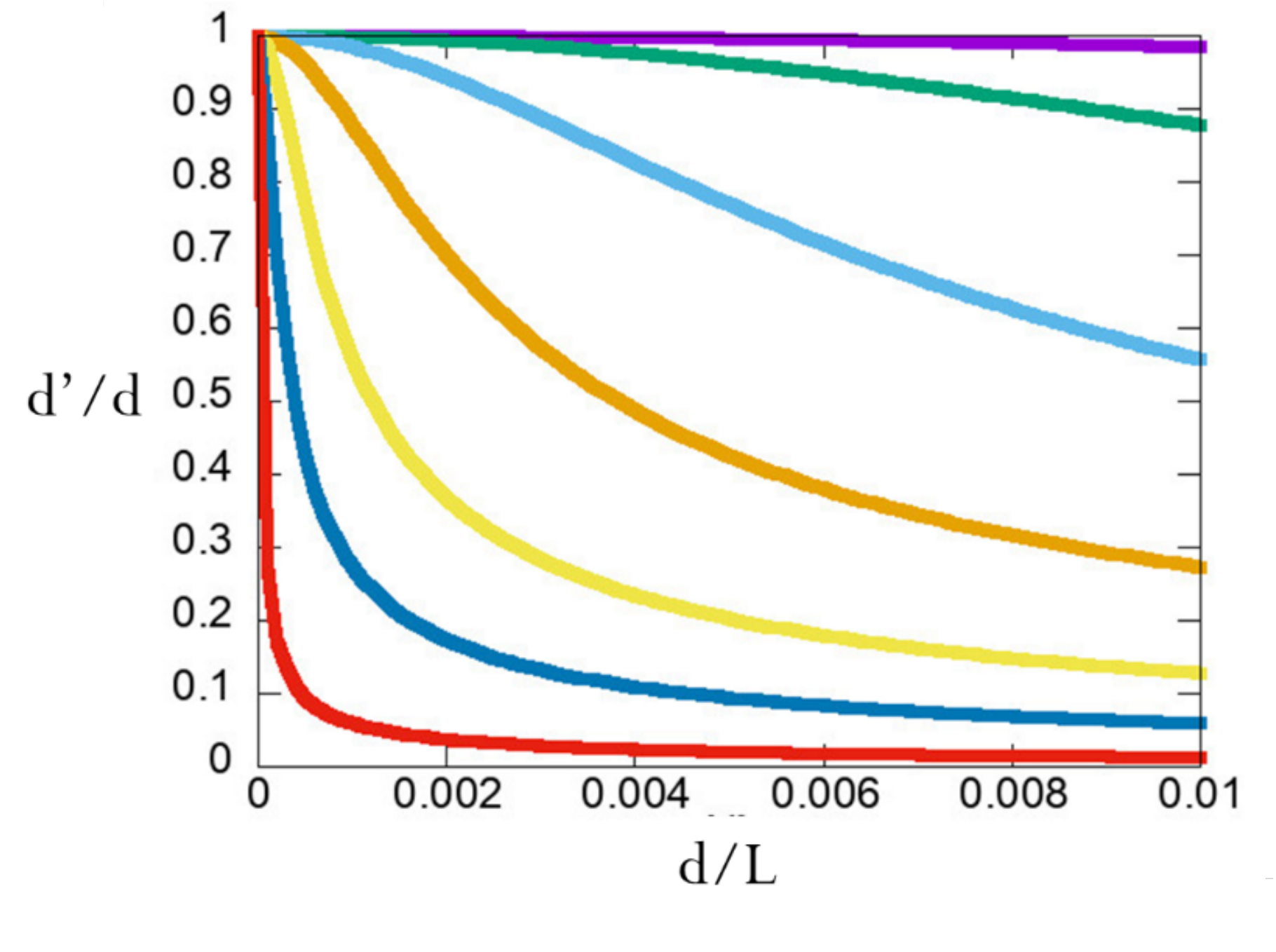}
\caption{The ratio $d'/ d$ as a function of the reduced separation $d/L$ for $n=10^k$ and with $k$ in the range from $1$ to $8$ (from top to bottom). The rapid drop of the ratio $d'/d$ and subsequent saturation at increasing values of $n$ is clearly apparent. Note that $d/L$ is always well below $1$, as it should be within Casimir effect theory.}
\label{fig2}
\end{figure}

At this point it is interesting to notice that, based on Eq.(\ref{eq2}), the relationship between $d$ and $d'$ takes the form
\begin{equation}\label{eq7}
  \left(\frac{d'}{d}\right)^3=\frac{1}{1+\frac{d^2}{d^2_n}},
 \end{equation}
with
\begin{equation}\label{eq8}
  d_n=\sqrt{\frac{C}{2\pi}}\frac{L}{n^{1/2}}\sim\frac{L/7}{n^{1/2}}.
\end{equation}
Equation (\ref{eq7}) can be also cast in a more compact for as follows:
\begin{equation}\label{eq9}
  \frac{d'}{d}=\frac{1}{\left(1+n\frac{d^2}{l^2}\right)^{1/3}},
\end{equation}
where we have set $l\sim L/7$. The expressions (7-9) constitute the cornerstone of the {\it Gedankenexperiment}, since they lead to a number of interesting considerations. The first one is that it shows the ratio $d'/d$ as a function of the reduced separation $d/L$, depending parametrically on the wavenumber $n$. This dependence is plotted in Fig.\ref{fig2} for different values of $n$. The graph indicates the existence of a characteristic wavenumber for each given value of the ratio $d/L$ below which the contraction of the space between the plates is really small. When approximating to such characteristic value of $n$, the ratio $d'/d$ experiments a fast decrease, followed by a very slow decay with $d/L$ which is basically undetectable in the relevant regime $d/L\ll 1$. The second one is the emergence of two regimes which for convenience are identified here as ``slow-mode'' and ``fast mode'' which are subsequently analyzed in the following sections.

\section{Fast and slow-mode regimes}

Based on Eq.(\ref{eq9}), two regimes can be identified: i) ``slow-mode'', characterized by
\begin{equation}\label{eq10}
n\frac{d^2}{l^2}\ll1,  
  \end{equation}
and ii) ``fast mode'', characterized by the opposite inequality. The slow-mode inequality given by Eq.(\ref{eq10}) implies that at each given value of the reduced separation $d/L$ there is a maximum wavenumber, which we call $n_1$, such that the ratio $d/d'$ is slightly below one, therefore corresponding to small contractions between the plates. By definition then: 
\begin{equation}\label{eq11}
  n_1=\frac{l^2}{d^2}.
\end{equation}
In the slow-mode regime $n\ll n_1$, Eq.(\ref{eq9}) reduces to:
\begin{equation}\label{eq12}
\frac{d'}{d}=1-\frac{n}{3}\frac{d^2}{l^2}=1-\frac{1}{3}\frac{n}{n_1},
\end{equation}
which, by definition, is just slightly smaller than one.

In the fast-mode regime, Eq.(\ref{eq9}) leads to:
\begin{equation}\label{eq13}
  \frac{d'}{d}\sim n^{-1/3}\left(\frac{l}{d}\right)^{2/3}=\left(\frac{n}{n_1}\right)^{1/3}.
\end{equation}
This expression shows that $d'$ decreases very slowly with the wavenumber $n$ and cannot reach the value zero for any finite value of $n$. In other words, the separation between the plates cannot be zero. We shall return to this point in the final part of this paper, when discussing the implication of the Eq.(\ref{eq13}) in the limit of the Planck scale.

At this point, it is worth noting that as required by the Casimir effect theoretical formalism, the ratio $d/L$ must be necessarily much smaller than one for Eq.(\ref{eq1}) to hold. Hence, as an example, we consider a ratio $d/L=10^{-3}$ which thus gives $n_1\sim 10^4$. Based on the Eq.(\ref{eq9}) and Eq.(\ref{eq11}), Eq.(\ref{eq6}) leads to
\begin{equation}\label{eq14}
\Delta t=\frac{d'}{nc}(1+10^{-4}n)^{1/3}.
\end{equation}

As it indicates, the quantity in brackets only becomes significant for wave numbers $n>10^4$. It should be noted that, considering the ``slow mode regime'' with $d=1nm$ and $c\sim 10^8$ $m/s$, it gives $d/c$ of the order of the attosecond ($10^{-18}s$). This means that only modes with $n \ll 10^4$ could be possibly detected within the current frontier of ultrafast technology. This is analyzed further in the following sections.

\subsection{Analysis of the slow-mode regime}
Considering $L/d \sim 10^3$, $n_1\sim 10^4$, hence for $n<n_1$ (``slow modes''), expanding the Eq.(\ref{eq12}) leads to:
\begin{equation}\label{eq15}
  \frac{d'-d}{d}\simeq -\frac{n}{3}10^{-4},
\end{equation}
which indicates a relative contraction in the order of the $4th$ digit for $n\sim 10$, a measurable effect as long such measurement is performed within the limits of the fluctuation lifetime $\Delta t$. Since, in this regime, $d'\simeq d$, one has 
\begin{equation}
\Delta t\sim \frac{d}{cn}.
\end{equation}
Given that the maximum value of $n$ in the slow mode regime is $n_1=10^4$, and approximating $c\sim 10^8 m/s$, we obtain a measuring time of $10^{-21}s$, which is between two and three orders of magnitude beyond the current technological possibilities even for measurement of ultrafast processes \cite{Grundmann}.

\subsection{Analysis of the fast-mode regime}

In the fast-mode regime, Eq.(\ref{eq13}) leads to
\begin{equation}\label{eq19}
  \frac{d'}{d}\simeq \left(\frac{10^4}{n}\right)^{1/3}\simeq 10 n^{-1/3}.
\end{equation}

This shows that, due to the small exponent $-1/3$, it takes very large wavenumbers $n$ to achieve substantial contractions. For instance, for $n=10^6$ a contraction is obtained of $d'=d/10$, and for $n=10^9$, it gives $d'=d/100$. Note that the limit $n\rightarrow \infty$ is unphysical since the wavelength of the photon cannot be lower than the Planck length $L_p$. This is discussed more in detail in the next paragraph.  Moreover, as it is readily seen from Eq.(\ref{eq14}), the corresponding fluctuation lifetime $\Delta t$ is of the order of $10^{-24}$ and $10^{-27}$ s respectively, hence far outside reach of the present technology. 

In summary, we have seen that a ``vacuum fluctuation'' between the plates, both of a high energetic mode ($n>10^4$) or a low energetic mode ($n<10^4$), produces a separation effect between the plates that is beyond experimental reach with today's technological capability. Perhaps a challenge is left for future experimental instrumentation developments. The following observation is relevant at this point. The single photon assumed to be extracted from within the plates, being a single mode, has an energy which does not grow with the transverse area $L^2$ of the plates, while the energy in between the plates does (see Eq.(\ref{eq1})). Accordingly, in order to compensate for this different scaling and in principle obtaining a measurable effect for $\Delta E$ and, therefore, on the distance variation between them, a highly energetic mode is needed, which then becomes unmeasurable within the associated extremely short lifetime $\Delta t$ of its corresponding fluctuation. On the other hand, assuming a photon with reasonable energies will eventually result in an undetectable displacement of the walls. 

\section{Down to the Planck scale}

In this section, the authors would like to take this {\it Gedankenexperiment} further and playfully speculate about the results obtained. In order to explore this matter, let us introduce the Planck wavenumber $n_P$ , as the value of $n$ such $\Delta t=t_P$ and $\lambda=L_P$, meaning the Planck time and the Planck wavelength respectively, $L_P$ being the Planck length.  By definition, we have:
\begin{equation}\label{eq20}
 n_P=\frac{2d}{L_P}.
 \end{equation}
The Eq.(\ref{eq13}) becomes minimum assuming the largest possible value for $n$, given by $n_P$. Using the definitions given by Eq.(\ref{eq20}) and Eq.(\ref{eq11}), we obtain:
\begin{equation}\label{eq21}
  d_{min}=d\left(\frac{n_1}{n_P}\right)^{1/3}\sim L^{2/3}L_P^{1/3}.
\end{equation}
This shows that even upon considering the most energetic possible mode (a Planckian photon), ``Casimir space'' cannot contract below $(L/L_P)^{2/3}$ Planck lengths. This result is intriguing, for it singles out a minimum length scale well above the Planck length.  Thus, the authors dare speculating that, although with an exchange of exponents between $L$ and $L_P$, an intriguing analogy holds with the holographic expression $L_{min}= L_P^{2/3} L^{1/3}$ . As is well known, this expression derives from assuming that the relevant quantum gravitational degrees of freedom associated with a given spatial region are located on the surface and not on the volume \cite{Stephens,Susskind}. The relation \ref{eq21} might also suggest a dependence of the boundary conditions of our {\it Gedankenexperiment.}

Coming back to the thought experiment and starting with Eq.(\ref{eq14}),
\begin{equation}
\Delta t\sim \frac{1}{nc}d'(1+10^{-4}n)^{1/3},
\end{equation}
when considering $n=n_P$, this approximates to
\begin{equation}\label{eq23}
  \Delta t\sim \frac{d'}{c}\frac{1}{10n_P^{2/3}}.
\end{equation}
By taking into account the definition of $n_P$, it leads to:
\begin{equation}\label{eq24}
  \Delta t_P\sim \frac{d'}{10c}\left(\frac{L_P}{d}\right)^{2/3}.
 \end{equation}
By considering Eq.(\ref{eq13}) and after some algebra, it is possible to arrive, as expected, to:
\begin{equation}
 \Delta t_P\sim \frac{n_1^{1/3}}{c}L_P\sim \frac{L_P}{c}\sim t_P.
\end{equation}
Note that $\Delta t_P$ as computed above is much smaller than the smallest time interval supported by the hypothetical time allowed by the Casimir theoretical formulation
\begin{equation}
  t_{min}=\frac{d_{min}}{c}=\left(\frac{L}{c}\right)^{2/3}t_P^{1/3}.
 \end{equation}
This might suggest some connection with a generalized Heisenberg principle and more particularly given the case that, in our {\it Gendankenexperiment}, the distance between the plates cannot be made arbitrarily small. More concretely smaller than the Planck length \cite{Kempf}. As a matter of fact, it is interesting, as a final consideration of this note, relating the energy between the plates with respect to the Planck energy, $E_P$. Starting with Eq.(\ref{eq1}) and taking into account Eq.(\ref{eq21}), we obtain:
\begin{equation}\label{eq26}
E_{min}\simeq -\frac{\hbar cL^2}{d^3_{min}}=-\frac{\hbar c}{L_P}=-E_P,
\end{equation}
where we have assumed $|E(d)|\ll E_P$ as it is appropriate for $d>d_{min}$. This result would seem paradoxical because, considering a fluctuation of the order of the Planck energy, would imply that $E=-2E_p$, which cannot be since $E$ is a free parameter. This apparent paradox is explained by considering that in fact $d'$ approximates only asymptotically to $d_{min}$, never to reach exactly this limiting value as expressed in Eq.(\ref{eq21}). More concretely, by taking Eq.(\ref{eq2}) with $C=1$, we compute:
\begin{equation}
E(d=L_P)=\frac{\hbar cL^2}{L_P^3}=E_P\left(\frac{L}{L_P}\right)^2\gg E_P.
\end{equation}

As this expression suggests, the apparent paradox mentioned above results from the fact the energy between the plates is singular as $d$ approaches zero, hence it takes much more energy than $E_P$ to push $d'$ from $d_{min}$ down to $L_P$. The interesting conclusion is that while $E(d')$ closely approaches $E(d_{min})$, $d_{min}$ still remains far above the Planck length.

\section{Conclusions}

Our excursion into the Casimir sandbox through the thought experiment proposed in this note leads to the following observations:

\begin{enumerate}

\item Considering the occurrence of a hypothetical ``vacuum fluctuation'' between the plates subject to the energy-time uncertainty leads to two differentiated regimes ruled by the scaling relationship $L/d=10^3$ between the lateral dimensions of the plates $L$ and their distance $d$, characterized by the value of the mode wave number $n$. One which we will call ``fast mode'' where $n>10^4$ and another one which we will call ``slow mode'' where $n<10^4$.

\item The potential shrinking effect between the plates of both, the high energetic mode ($n>10^4$) or the low energetic mode ($n<10^4$), cannot be measured with today's technological capability. Perhaps, in the low energetic mode, future advances in technology might be able to cope with the challenge.

\item By considering a fluctuation at the order of the Planck energy, the momentarily contracted distance between the plates cannot reach a value $d_{min}$ below $(L/L_P)^{2/3}$ Planck lengths. More precisely, $d_{min} = L^{2/3} L_P^{1/3}$, $L$ being the lateral extension of the plates and $L_P$ the Planck distance. The authors noticed the intriguing analogy between the expression for $d_{min}$ and the one $L_{min}= L_P^{2/3} L^{1/3}$ which derives from assuming that the relevant degrees of freedom or quantum gravity are located on the surface and not on the volume (``holographic principle'').

\item  If the diminished distance $d'$ between the plates, following the thought experiment, approaches the minimum distance $d_{min}$, it does it asymptotically. This leads to the interesting observation that while the energy between the plates, $E(d')$, approaches the minimum one, $E(d_{min})= E_P$, the minimum Casimir distance $d_{min}$ remains far above the Planck length, $L_P$. Conversely, the Casimir energy at the Planck scale exceeds the Planck energy by a factor $(L/L_P)^{2/3}$.

\item One may finally wonder whether the present analysis would hold if the geometry of the conducting plates is not flat. Although this is beyond the scope of this manuscript, it is interesting to note that modifying the shape of the plates can turn the sign of the energy from negative to positive, thus leading to a repulsive rather than an attractive interaction \cite{wilczek}. This has been shown, for example, in Ref. \cite{PRL_LEVIN2010}, where a small elongated metal particle in vacuum is subject to a repulsive force when centered above a metal plate with a hole. Repulsive long-range forces, of quantum electrodynamic origin, have been also measured between materials with suitable optical properties and immersed in a fluid \cite{capasso_nature}.
Even more intriguing is the realm of soft condensed matter, where Casimir-like forces have been found to act between surfaces immersed in a binary fluid close to its critical point. Such forces, caused by the fluctuations of the concentration (whose relevant scale is $k_BT$, where $k_B$ is the Boltzmann constant and $T$ the temperature) within the fluid film separating the surfaces, can be either attractive or repulsive, depending on the adsorption preference of the fluid in contact with the solid body \cite{gambassi_nature,gambassi_iop}. Finally, it is known that near-contact forces between macroscopic bodies play a major role on the rheology of soft materials \cite{montessori}.
\end{enumerate}

\section*{Acknowledgments}

All authors warmly acknowledge Andrea Gambassi for useful discussions and a critical reading of the manuscript. SS and AT would like to acknowledge funding from the European Research Council under the Horizon 2020 Programme Grant Agreement n. 739964 (``COPMAT''). This document has the CERN Open reference CERN-OPEN-2021-002.


\begin{thebibliography}{99}

\bibitem{Wheeler} J. A. Wheeler, On the nature of quantum geometrodynamics, Annals of Physics, Volume 2, Issue 6, Pages 604-614 (1957). 
\bibitem{Casimir} H. B. G. Casimir, On the attraction between two perfectly conducting plates, Proc. Kon. Ned. Akad. Wet. 51, 793 (1948).
\bibitem{Jaffe} R. L. Jaffe, Casimir effect and the quantum vacuum, Phys. Rev. D, 72, 021301(R) (2005).
\bibitem{Grundmann} S. Grundmann, Zeptosecond birth time delay in molecular photoionization, Science, Vol. 370, Issue 6514, pp. 339-341 (2020).
\bibitem{Stephens} C.R. Stephens, G. 't Hooft and B.F. Whiting, Black hole evaporation without information loss, Classical and Quantum Gravity, 11 (3), 621-648 (1994);
\bibitem{Susskind} L. Susskind, The World as a Hologram, Journal of Mathematical Physics, 36 (11): 6377-6396, (1995).
\bibitem{Kempf} A. Kempf et al., Hilbert space representation of the minimal length uncertainty relation, Phys. Rev. D, Vol.52, 1108-1118, (1995).
\bibitem{wilczek} Q.-D. Jiang, F. Wilczek, Chiral Casimir Forces: Repulsive, Enhanced, Tunable, Phys. Rev. B 99, 125403 (2019).
\bibitem{PRL_LEVIN2010} M. Levin, A. P. McCauley, A. W. Rodriguez, M. T. H. Reid, and S. G. Johnson, Phys. Rev. Lett. 105, 090403 (2010).
\bibitem{capasso_nature} J. N. Munday, F. Capasso, V. A. Parsegian, Measured long-range repulsive Casimir-Lifshitz forces, Nature 457, 170-173 (2009).
\bibitem{gambassi_nature} C. Hertlein, L. Helden, A. Gambassi, S. Dietrich, C. Bechinger, Direct measurement of critical Casimir forces, Nature 451, 172-175 (2008).
\bibitem{gambassi_iop} A. Gambassi, The Casimir effect: From quantum to critical fluctuations, J. Phys.: Conf. Ser 161, 012037 (2009).
\bibitem{montessori} A. Montessori, M. Lauricella, N. Tirelli, S. Succi, Mesoscale modelling of near-contact interactions for complex flowing interfaces, Journ. Fluid. Mech. 872, 327-347 (2019).
\end{thebibliography}
\end{document}